# Resistive and magnetoresistive properties of $CrO_2$ pressed powders with different types of inter-granular dielectric layers.

Dalakova N.V.[1] Belevtsev B.I.[1], Beliayev E.Yu.[1], Bludov O.M.[1], Pashchenko V.A.[1], Osmolovsky M.D.[2], Osmolovskaya O.M.[2]

[1] Institute for Low Temperature Physics and Engineering, National Academy of Sciences of Ukraine, Kharkov 61103, Ukraine, E-mail: dalakova@ilt.kharkov.ua
[2] St. Petersburg State University, Department of Chemistry, St. Petersburg, 198504, Russia

Resistive, magnetoresistive and magnetic properties of four kinds of pressed $CrO_2$ powders, synthesized by hydrothermal method of chromic anhydride have been investigated. The particles in powders constituted of rounded particles (diameter 120 nm) or needle-shaped crystals with an average diameter of 22.9 nm and average length of 302 nm. All of the particles had a surface dielectric shell of varying thickness and different types (such as oxyhydroxide $\beta$-CrOOH or chromium oxide $Cr_2O_3$). For all the samples at low temperatures we found non-metallic temperature dependence of resistivity and giant negative magnetoresistance (MR). The maximum value of MR at low temperatures (T ≈ 5 K) is ≈ 37% in relatively small fields (0.5 T). At higher temperatures there was a rapid decrease of MR (up to ≈ 1% / T at T ≈ 200 K). The main objective of this work was studying the influence of properties and thickness of the intergranular dielectric layers, as well as $CrO_2$ particle shape, on the magnitude of the tunneling resistance and MR of the pressed powder. The new results obtained in this study include: (1) detection at low temperatures in powders with needle-like particles a new type of MR hysteresis, and nonmonotonic MR behaviour with increasing magnetic field (absolute value of the MR at first grows rather rapidly with the field, and then begins diminishing markedly, forming a maximum), and (2) detection of non-monotonic temperature dependence $H_p(T)$, where $H_p$ - a field in which the resistance in a magnetic field has a maximum, as well as finding discrepancies in values of $H_p$ and coercivity fields $H_c$, (3) detection of the anisotropy of MR, depending on the relative orientation of the transport current and the magnetic field, (4) a new method of synthesis, to regulate the thickness of dielectric coating.

## 1. INTRODUCTION

Chromium dioxide ($CrO_2$) is a ferromagnet with the Curie temperature $T_C$ ≈ 390 K. For a long time, this material (in the form of fine-grained powder) has been widely used for magnetic recording. The interest in chromium dioxide (in addition to the practical importance) is due to the fact that it is a half-metal [1-3]. In such metals, the electrons in a conduction band at the Fermi level has only one spin polarization. At sufficiently low temperatures, the spin polarization in $CrO_2$ can approach 100% [2]. Intrinsic magnetoresistance of CrO2 single-crystal is about 1%/T at room temperature [3]. In the case of pressed $CrO_2$ powder with particles coated by a thin dielectric layer (which is a composite granular material), MR become giant, reaching more than 30% at low temperatures in small fields [1, 4]. This may be of interest for practical applications. In this case, MR is caused by the presence of granular structure. In this material the dielectric layers prevent direct ferromagnetic exchange between the neighboring grains, permitting, however, the intergranular tunneling of electrons. Thus MR of the granular material is extrinsic. The probability of electron tunneling depends on the relative orientation of the magnetization vector in neighboring grains [3], and is very sensitive to the applied magnetic field. It has maximum when the moments in the adjacent grains are oriented parallel to each other [3]. In the literature, this kind of electron tunneling is called spin-dependent, and the corresponding magnetoresistance is called tunneling magnetoresistance. It's value depends on the thickness and dielectric properties of intergranular layers, which are largely determined by the technology of preparation of pressed powder.

In work [5] (with some of the authors of this paper) the samples, consisting of rounded (diameter about 120 nm) particles of $CrO_2$ with dielectric coating (~ 1 nm) of oxyhydroxide-$\beta$-CrOOH were prepared and analyzed. In this paper the resistive and magnetoresistive properties of compacted powders of other species, consisting of needle-shaped particles of $CrO_2$ were investigated. In this case, we had the dielectric coatings of another kind, in particular, consisted entirely of oxide $Cr_2O_3$. The influence of the dielectric properties of the barriers between ferromagnets, including the properties of interfaces between a ferromagnet-insulator (and the role of structural disorder in the barrier) on the tunnel MR is currently one of the most important and insufficiently investigated problem of ferromagnetic tunnel junctions [3, 6]. In the present paper in the framework of this problem we study the influence of the properties and thickness of intergranular dielectric layers, as well as particle shape, on the magnitude of the tunneling resistance and MR of pressed $CrO_2$ powders.

## 2. METHOD OF PREPARATION AND INVESTIGATION OF SAMPLES

The synthesis of chromium dioxide in present study was carried out by hydrothermal method. General features of the used technologies are described in [7]. As the main stage, the synthesis of chromium dioxide from the mixture of $CrO_3$, water and special additives, number and type of which determines the nucleation, growth, size and shape of particles [5, 7]. The synthesis was carried out in an autoclave at the pressure 32 MPa and temperature not exceeding 330 C°. In such ways three types of pressed chromium dioxide powders with needle-shaped particles were prepared and characterized.

The powder number 1 was obtained after the above hydrothermal synthesis and drying at 150 C°, followed by sieving through a sieve and subsequent pressing. The sample contained a germ-phase of $Cr_2MoO_6$ heteroepitaxially covered by $CrSbO_4$. The ratio of elements (atomic = mol) in powdered (molybdenum : antimony : chrome) mixture was (0.2 : 2.0 : 100). Consequently, the powder had about 6% by weight of non-magnetic germinal phases. The role of additives in the synthesis - the creation of germ particles of completely different composition, but with the structure like chromium dioxide, and with lattice parameters different from those of chromium dioxide up to 5%. In the presence these additives, the chromium dioxide particles formed from the aforesaid solution (without any admixture of "intermediate" chromium oxides) directly on the alien germs – the process of heteroepitaxial crystallization took place. These embryos were created due to the formation of mixed oxides with a rutile (or three-rutil)-type structure that contained chromium (3) and/or molybdenum (6), etc. These phases were formed at 140 – 200 C° and could be consistently built one on another, bringing the germ to the optimum size. The chromium dioxide (on these germs) begins to form at 220-230 C° and ends at 320 – 350 C°, depending on the amount of water. In the end of synthesis in the autoclave we have chromium dioxide, which includes the nucleation particles and the water in the form of vapor and liquid.

The resulting needle-like particles were coated with a naturally degraded layer consisting of a mixture of amorphous $\beta$-CrOOH and some amount of occluded chromic acid. Both compounds formed by the reaction of chromium dioxide and water vapor during the cooling of the product in an autoclave. Degraded layer, although it is quite friable, allows tunneling of electrons. The powder number 1 served as a starting material for the preparation of other powders by thermal and chemical treatment.

Further processing of powders was carried out as follows. Part of the resulting powder number 1 was heated in air at 320 C°, with the result that the content of $\beta$-CrOOH and chromic acid dramatically decreased due to their interaction with one another leading to formation of $CrO_2$ and due to the oxidation of $\beta$-CrOOH to $CrO_2$ by oxhatygen from the air. In the end, the surface layer consisted almost exclusively of $CrO_2$ (this stage is called the "enrichment process"). Another part of this powder was treated with a solution of a reducing agent [5], resulting in the particles coated with a thick stabilizing layer of crystalline orthorhombic β-CrOOH (stage of the stabilization). The stage of stabilization, due to the changes in composition of the surface layer leads to formation on the granuli surface a dense, well protecting from the external environment, dielectric shell which has no pores.

Calcination in a flow of helium at T = 270 C° turns the oxyhydroxide (β-CrOOH) shell into oxide one ($Cr_2O_3$). This procedure was applied to the powder, after a number of stages of enrichment and stabilization. The resulting material was named the powder number 2. The same treatment was applied to the number 1 powder to obtain the powder number 3. The difference from the powder number 2 was the use of a reducing solution with higher concentrations of the components, to provide formation of thicker reduced layer. Both powder (№ 2 and № 3) had a thin surface layer of $Cr_2O_3$ on the particles, but for the powder number 2 the dielectric layer was thinner. The thinnest dielectric layer had the powder number 1. $Cr_2O_3$ that has a rhombohedral lattice, is an antiferromagnet with a Néel temperature 309 K, and it exhibits a weak magnetoelectric effect.

The samples having the shape of a parallelepipeds with dimensions of 3×5×12 $mm^3$ were made by cold pressing the powders of needle-shaped particles. The density of our pressed pellets was approximately 40% of the X-ray density of the material. In the known literature [5] needle-shaped $CrO_2$ powders were pressed up to density of 60%. Such a strong compression could be done when the crystallites had a weakly expressed anisotropy of the form. It should also be noted that the particles having a needle-like shape, should orient themselves in the plane being perpendicular to the direction of compression when being compacted in tablets, although their mutual orientation in the planes could be quite messy. And so, there may be some differences of transport properties for measuring currents parallel and perpendicular to the direction of pressing.

The average diameter of needle-shaped particles for the samples № 1-3 was 22.9 nm and the average length 302 nm. The average particle size was determined by electron micrographs. This is usually done for 30 particles. Then the arithmetic mean size was calculated. The scatter of readings was not more than 4-5%. For example, for an average diameter of the needle-like particles we obtained the value 22.9 ± 0.8 nm. Sometimes the measurements were made for a larger number of particles (up to 400). But the difference in data compared with 30 particles still did not exceed 5%. Along with the samples with needle-shaped particles the sample number 4, consisting of rounded (diameter about 120 nm) particles of $CrO_2$ with dielectric layers of stabilizing oxyhydroxide β-CrOOH has also been investigated. The method of its preparation is described in [5]. The density of this compacted sample was much higher (about 60% of the X-ray density).

The average thicknesses of the dielectric layers for the samples studied were different (for example, for the series of samples with needle particles (№ 1-3) the layer's thickness have increased with the number of sample). For all samples, the thicknesses of layers were of the order of 1 nm, although even slight differences were sufficient to significantly impact on the tunneling resistance. Evaluation of thickness of dielectric layers was determined both directly (via high resolution TEM), and indirectly, judging from the consumption of reagents in the formation of dielectric layers or the magnitude of specific magnetization. By these methods, we determined the thickness of the dielectric layer for samples number 3 and number 4 (2.1 nm and 3.6 nm, respectively).

The powders were characterized by electron microscopy, X-ray and magnetic studies. Micrograph of needle powder, obtained in a transmission electron microscope, is shown in Fig.1. The lattice parameters obtained for needle particles (a = 0.4424 nm, c = 0.2916 nm for rutile-type lattice) correspond to the known data for pure $CrO_2$ [2]. Magnetic properties were measured on a vibration (77 Hz) and a SQUID (Quantum Design) magnetometers. An example for the magnetization hysteresis curve is shown in Fig.2. The Curie temperature of the samples in a small field was about 390 K. The temperature and magnetic field dependence of magnetization revealed interesting features that will be presented in a separate publication. The main characteristics of the samples listed in Table 1.

Resistance measurements were made by standard four-point probe technique at a given current (J = 100 mA). The measurements have been carried out at this low current at which Ohm's law is obeyed. The distance between the potential contacts was 8 mm. Magnetic field dependences of resistance were recorded in the temperature range (4.4 ÷ 200 K) in fields up to 1.5 Tesla. Magnetoresistive effects in the present work were studied at a magnetic field directed perpendicular to the current. Separate series of measurements were made in a magnetic field parallel to the transport current that will be discussed below. Protocol of MR measurements corresponded to the usual protocol for measurement of magnetization hysteresis curves. MR measurements were performed after entering the maximum field H=1.5 T.

### 3. RESULTS

Fig.3 shows the temperature dependence of resistivity of the samples. Numbers shown at curves in this figure correspond to numbers of samples described in the previous section and Table 1. The most resistive sample is the sample number 4 with a stabilized dense shell of oxyhydroxide for the $CrO_2$ particles of a rounded shape. This is not surprising, since this sample has the greatest thickness of intergranular layers (3.6 nm). In this sample, the temperature dependence of the resistance is close to exponential [ $\rho(T) \propto \exp(1/T)$ ] for $T < 20$ K. Above 20 K the $R(T)$ dependence is not so steep. For samples with needle-shaped $CrO_2$ particles $\rho(T)$ dependence for $T \leq 50$ K corresponds to the Mott's law of variable range hopping conductivity for 3D systems: $\rho \approx \rho_0 \exp(T_0/T)^{1/4}$. For $T > 50$ K there is some deviation from the Mott's law.

For the samples with needle particles the largest resistivity has the sample №3 with a thick dielectric layer of $Cr_2O_3$ (2.1 nm), and the smallest – the sample №1 with stabilized dielectric coating round the particles. With increasing temperature the sample №1 has a minimum of resistance at $T_{\min} \approx 140$ K, and a transition to the metallic temperature dependence of resistivity ($d\rho/dT > 0$) takes place. Such cases sometimes are referred to as the insulator-metal transition with increasing temperature. The lower values of the resistivity for this sample (as to compare with samples №2 and №3) may be related to the inhomogeneity of the thickness and possible local discontinuity of dielectric shell round the particles. This kind of resistance minimum is a fairly typical phenomenon in polycrystalline or granular oxides of transition metals with a non-uniform thickness of the dielectric layers. The main reason for it is the transition from activated to non-activated electron tunneling for temperatures $T > T_{\min} \approx 140\,\text{K}$ and formation of a percolation channel consisting of series of metal granules with weak non-activated tunneling barriers or simply metallic "short-circuits". In details this mechanism of minimum resistance in granular magnetic metals is described in [8]. There are also some models of this type of minimal resistance, see, e.g., [9].

Fig.4 shows the magnetic-field dependence of MR, $[R(H) - R(0)]/R(0)$, for the sample №4 with round-shaped $CrO_2$ particles recorded at $T$ = 4.17 K and 20.43 K. The lower panel of the figure (T = 20.43 K) shows typical of polycrystalline oxides of transition metals magnetic hysteresis behavior of the tunnel MR, which is fully consistent with the magnetization hysteresis (see explanation in [3]). In this type of hysteresis at low fields, there are two peaks of positive MR at characteristic fields $H$, equal to $+H_p$ and $-H_p$, where the value of $H_p$ should correspond to the coercive force $H_c$ [3]. This usual form of hysteresis was observed for the MR of sample №4 only at sufficiently high temperatures ($T \geq 15\,\text{K}$). At low temperatures, the form of hysteresis become more complicated (see Fig.4, hysteresis of MR at $T$ = 4.17 K). In this case, along with two positive peaks of MR, there was an additional crossing of MR curves for the forward and backward sweep of the magnetic field in fields slightly above $H_p$. A possible reason for this hysteresis is discussed in [5] and is associated with the percolation nature of the tunneling conductance of the granular system at low temperatures where the conductivity of the entire system can be determined by few percolation current channels. In this case, the hysteresis in the fields above $H_\text{p}$ is associated with switching of small amount of percolation current channels for increasing and decreasing field at low temperatures.

Despite the rather large thickness of intergrain layers (3.6 nm), the highest among the measured samples, negative MR for the sample №4 at low temperatures is high enough (over 18%). It should be noted that according to the known literature the dependence of MR on the layer thickness has almost not been investigated, but in spite of this some considerations in this regard can be made. One can be confident enough to believe that MR increases with the thickness of dielectric layer, but to a certain limit. For sufficiently thick insulating intergranular layers granular samples are transformed into the system of completely isolated granules, in which tunneling and, accordingly, tunnel MR is absent completely.

In hysteresis behavior of MR for samples with needle-like $CrO_2$ particles significant differences from the hysteresis of the sample №4 with a rounded shape of the particles have been found. As an example, these differences can be seen in the sample №2 (Fig.5). It can be seen from this figure that the hysteresis increases with decreasing temperature. MR hysteresis at $T = 20$ K and 10 K completely corresponds to the types of hysteresis behavior described above for the sample №4 (Fig.4). But hysteresis for $T \leq 5$ K has, however, a number of features that not previously been mentioned in known literature not only for pressed $CrO_2$ powders, but also for other granular magnetic metals. This (third) type of hysteresis is shown in Figure 5 at T = 4.4 K. In this case, in small fields instead of two peaks of positive MR at $|H| = H_p$ there observed a more complex structure of $R(H)$ curves, which will be discussed in a separate publication. In addition, for the sample №2 at T = 4.4 K (Fig. 5) a nonmonotonic dependence of MR with increasing magnetic field is observed: the absolute value of MR at first rapidly increases with the field, and then begins to decrease significantly, forming a maximum. This behavior at low temperatures was found for all three samples (№ 1, № 2 and № 3) with needle particles (for the sample №3, this behavior was most pronounced, but for the sample №1, it manifested itself rather weakly). This contradicts the commonly observed MR hysteresis behavior for magnetic oxides of transition metals. For these materials with increasing field there usually observed monotonic increase of the negative MR: at first a rather sharp increase in low fields, followed by a much slower increase in higher fields. It is believed [3] that it reflects the field dependence of magnetization (Fig.2): strong growth in low fields, followed by a weak increase in high fields. A possible reason for the inconsistency of the magnetization hysteresis and the hysteresis of MR in inhomogeneous granular magnetic systems was shown earlier in [5]. The magnetization measured by a magnetometer, is determined by the contribution of all the granules of the system. At the same time in a percolation system with unequal intergranular tunneling barriers conductivity is determined by the presence of "optimal" chains of grains with a maximum tunneling probability [10]. In conditions of activated conduction, which take place for the samples studied (Fig.3), the number of conducting chains decreases with decreasing temperature [10] and with increasing magnetic field [11], so that at sufficiently low temperatures and high magnetic fields, percolation conductive network can be reduced to a single conducting channel [10,11]. Under these conditions, the conductivity at low temperatures may be determined by the small volume fraction of granules, and the local magnetic properties of this part of the granules may differ from the global behavior of the magnetization measured by a magnetometer. This may be one of the reasons of strong differences between values of $H_\text{c}$ and $H_\text{p}$ found in [5]. In the present work this phenomenon is observed for the pressed powders of acicular $CrO_2$ particles, which will be discussed below.

The temperature dependences of the field $H_\text{p}$ and the coercive force $H_\text{c}$ of sample №3 are shown in Fig.6. Typically, the value of $H_\text{c}$ for ferromagnets is the highest at low temperatures and decreases with increasing temperature (in the limit it goes to zero when approaching $T_\text{c}$). The behavior of sample №3 in Fig.6 corresponds to the expected one. At the same time dependence of $H_\text{p}(T)$ for samples with needle-like particles (Fig.6) is unusual. First, these samples did not satisfy the relation $H_p(T) \approx H_c(T)$, which is expected and usually observed for the pressed powders with a rather small (submicron) sizes, including $CrO_2$ powders [1, 4]. Second, the relationship was non-monotonic: in the range 50-100 K, the value of $H_p$ is much higher than $H_C$, and at higher temperatures the difference between $H_p$ and $H_c$ decreases significantly. Previously, non-monotonic dependence was observed for the pressed $CrO_2$ powders, consisting of rounded particles with a dielectric β-CrOOH coating [5] (these correspond to the powder sample №4, investigated in this paper). In these samples with particle size 120 nm one could expect that all the particles are single-domain, as it is well known [12] that the critical diameter of the spherical single-domain particles of $CrO_2$ is about 200 nm. The scatter of the size of the granules, as well as weak tunnel barriers between some grains leads, however, to the fact that a small portion of the granules are in a multidomain state. Meanwhile, as pointed out in [5], at low temperatures a few conductive percolation channels, which provide the main contribution to the conductivity, mainly consist of multidomain particles. In these particles, the local magnitude of $H_c$ is smaller than in single-domain particles, which leads to a decrease in measured quantities of $H_p$ at a sufficient temperature decrease.

The treatment in [5] is largely consistent with the results of the study [13] for pressed powders of manganites $La_{2/3}Sr_{1/3}MnO_3$ with different particle sizes. In this study it was found that for multidomain particles of manganite $H_p > H_c$ is satisfied. The difference between values $H_p$ and $H_c$ decreases with decreasing particle size, and with a sufficient reduction of particle size, they become single-domain, resulting in the ratio $H_p = H_c$. In the single-domain particles with increasing magnetic field the magnetic moments of grains tend to orient parallel to the applied field. This is done by uniform rotation of the magnetic moments of particles in a magnetic field. In this case there is a clear matching between the values of MR and the measured magnetization of the sample, so that the relation $H_p \approx H_c$ become true [3]. In multidomain particles magnetization reversal (i.e., the change in the direction of the magnetization vector to the opposite one) more easily occurs by nucleation and growth of the domain with the opposite direction of the magnetic moment [14, 15]. The motion of domain walls has almost no effect on MR caused by the spin-dependent magnetic tunneling [13]. In such circumstances, $H_p \neq H_c$ inequality holds and there is no correlation between the measured magnetization and MR. These concepts are also applicable to samples № 1-3 with needle-shape particles. Particles (with an average diameter of 22.9 nm and average length of 302 nm) are clearly multidomain. Many aspects of the relationship, $H_p$ and $H_c$ yet still remain unclear and require further research. In particular, remains largely unclear a non-monotonic dependence $H_p(T)$ found in the present study.

The data in Table 1 show that the value of $H_c$ for the powder sample №4 with particles of rounded shape is much smaller than for the samples № 1-3 with acicular particles. Sample №4 had a smaller shape anisotropy and lower porosity than the samples with numbers 1, 2 and 3. And the specific surface area for this sample is 2.9 times less. In addition, because the crystal lattice of oxyhydroxide (rutile) is only slightly different from the lattice of the basic material $CrO_2$, the interface between the surface of the particles and the dielectric shell has no strong lattice distortions, violating the magnetic structure of $CrO_2$. Thus, the sample №4 was more homogeneous magnetically and did not has large internal stresses impeding the processes of magnetization reversal.

According to Table 1, the value of MR for samples with acicular particles with dielectric layers of $Cr_2O_3$ is much higher than the MR of sample №4 with particles whose shape is nearly spherical, and intergranular layers consist of β-CrOOH. In needle-shape samples the value of MR increases with the average thickness of the intergranular layers of $Cr_2O_3$. In pressed powders, the dielectric layers do not have a strict uniform thickness, and accordingly there is some statistical and spatial variation of these quantities. As a consequence of this some of intergranular contacts are weak barriers and the current passage should not be activated tunneling. Availability of immediate electrical intergranular contacts ("shortcuts") is also possible. All this may weaken the role of spin-dependent tunneling and reduce the tunnel MR in the samples with sufficiently thin intergranular layers. It is known that, compared to the shells of β-CrOOH, dielectric shells of $Cr_2O_3$ are thermodynamically more stable and give more reliable protection of $CrO_2$ particles from the effects of environment. Therefore, MR for samples with intergranular layers of $Cr_2O_3$ is higher than the MR for samples with rather loose layers of β-CrOOH. It is also natural that the increase in the average thickness of the intergranular layers of $Cr_2O_3$ leads to strengthening of the tunneling nature of conductance and to increase in the tunnel MR (Table 1).

Another important result of this work is the discovery of the anisotropy of MR, depending on the relative orientation of the magnetic field and transport current. The MR data measured for perpendicular mutual orientation of field and current ($H \perp J$) are given in Table 1. In the parallel orientation ($H \parallel J$) MR was significantly higher. For sample №3 at $T \approx 5$ K in a field of 0.3 T MR $=[R(H)-R(0)]/R(0)$ is -36.3% at $H \perp J$ and 40% at $H \parallel J$. With the field growing the anisotropy of MR is being reduced. The dependence of MR on the mutual orientation of field and current is well known for 3d metals and is considered to be a characteristic (intrinsic) property of ferromagnets [16,17]. It is believed that this type of MR anisotropy is caused by the anisotropic spin-orbit interaction. For bulk ferromagnets the magnitude of this effect is usually about 1% [1]. In this paper, the anisotropy of MR, $[R(H_\parallel)-R(H_\perp)]/R(0)$, was significantly higher (4%). This rather large value does not enable us to refer it fully to the influence of spin-orbit interaction, in accordance with the models [16,17]. We can not exclude that there are contributions from other sources for this kind of anisotropy. It was stated above that in pressed tablets the needles are placed in the planes in a haphazard way. In parallel orientation ($H \parallel J$) magnetic field is directed along these planes, while in ($H \perp J$) the field is perpendicular to these planes. The difference in the transport properties for the measurements when currents are parallel and perpendicular to needle planes is quite possible, and this may contribute to the observed anisotropy of the MR. This issue deserves further investigation.

Tunnel MR of pressed $CrO_2$ powder decreases rapidly with increasing temperature above 5 K (Fig.7). For example, for sample №4 raising the temperature up to 180 K leads to reduction of MR by two orders of magnitude (0.2%). In previous papers [1, 4, 17], it was noted that the fall of MR with increasing temperature is proportional to $\exp(-T/T_m)$, i.e. proceed exponentially. The samples investigated in this study also follow this law, as it is evident from the semilogarithmic scale graphs in Fig.7. The values of $T_m$ are dependent on the range of magnetic field in which the measurements of MR were carried out [1, 4, 17]. The values of $T_m$ for the low fields near $H_p$ where a sharp decrease in resistance with increasing magnetic field takes place, are lower than in high fields, where the saturation of MR takes place. For the temperature dependence of MR, as shown in Fig.7, $T_m$ = 45 K for sample №4 and $T_m$ = 60 K for the acicular samples №2 and №3.

Tunnel MR of granular ferromagnets, including pressed $CrO_2$ powders, is determined only by direct tunneling of charge carriers with spin conservation. The processes of passage of the tunnel barriers, in which the spin of charge carriers is not preserved (for example, through spin reorientation on impurities and defects at the grain boundaries) do not contribute to the measured MR. In the known literature [1, 2, 3, 4, 18, 19, 20, 21], the main reason for the sharp fall in MR with increase in temperature is mainly associated with the decrease of spin polarization $P$ with increasing temperature. According to [2, 3, 20, 21] MR is related to the spin polarization by the formula

$$MR = \frac{P_e^2}{1+P_e^2} \quad (1)$$

where $P_e$- effective polarization for electron tunneling in the barrier without spin-flip. According to this formula, the maximum of MR for a magnetic tunneling could be 50%. Note that in our study, the maximum of MR was about 40%, which corresponds to $P_e \approx 82\%$. In [22], using the method of micro-contact spectroscopy, the values of spin polarization for $CrO_2$ were measured and obtained the value of $P = 90 \pm 3{,}6\%$ at $T = 1.6$ K. This is consistent with our result (82% for $T \approx 5$ K).

In [3] there was given the following expression for the tunneling MR:

$$\frac{\Delta\rho}{\rho_0} = -\frac{JP}{4kT}\frac{[M^2(H) - M^2(0)]}{M_s^2} \qquad (2)$$

where $J$ - the exchange interaction constant, $M_s$ - the saturation magnetization. It follows from (1) and (2) that the tunnel MR depends primarily on the polarization.

In the known literature, the discussion on reducing the tunneling resistance with increasing temperature is mainly qualitative. Some specific theoretical models can be found in [1, 3, 4, 6, 18, 19, 21]. In particular, in [21] presented a phenomenological model in which the authors suggest that the polarization $P$ is proportional to the magnetization of the surface layer of ferromagnetic particles, so that both these quantities decrease with increasing temperature under the influence of spin-wave excitations. For low temperatures range, the author of [21] gives the following expression for $P(T)$:

$$P(T) = P_0(1 - \alpha T^{3/2}) \qquad (3)$$

Where $P_0$ is the polarization at $T = 0$ K, α - the material dependent constant.

Another influencing factor in the fall of MR with increasing temperature is the existence of parallel channels for transmission of carriers through intergranular boundaries in which the spin is not preserved. Such processes include, for example, the mentioned above passage of grain boundaries with spin-flip on impurities and lattice defects, multistep tunneling and other processes [1, 4, 18, 19]. The role of the processes of the grain-boundary transmission without conservation of spin increases with increasing temperature, which also contributes to the rapid decrease of MR with increasing temperature.

One of the most interesting (and unexpected) results of this study is the discovery of non-monotonic dependence of MR under magnetic field increase (at first the absolute value of MR grows rather rapidly with the field, and then begins to markedly decrease, creating maximum) (see Fig.5 for sample №2 at T = 4.4 K). This effect has not been mentioned before in the known literature for granular ferromagnetic oxides. In our study, we observed this effect only at low temperatures ($T \leq 5$ K) for powders of needle-like particles, and the effect is amplified as the formation of dense intergranular layers of $Cr_2O_3$ proceeds. This is illustrated in Fig.8, which compares the MR curves for the samples №1 and №2. In the sample №1 acicular particles were coated by a naturally degraded layer consisting of a mixture of amorphous β-CrOOH and chromic acid. This layer is very loose and nonmonotonic change in MR with increasing magnetic field is manifested rather weakly. At the same time for sample №2 with thick shells of intergranular $Cr_2O_3$ non-monotonic field dependence of the MR becomes very pronounced .

Non-monotonic changes of MR with increasing of magnetic field means that the resistance of the sample at first decreases and then begins to increase, i.e. there is a positive contribution to MR. The origin of this positive MR may be that a sufficiently strong magnetic field changes the properties of the intergranular tunneling barriers, which consist of layers of $Cr_2O_3$. In the known theories of tunneling MR of granular ferromagnetic it is usually assumed that the dielectric barriers are nonmagnetic. However, $Cr_2O_3$ is an antiferromagnet, so that the magnetic field can change its properties. According to [23], dielectric $Cr_2O_3$ barriers between the $CrO_2$ particles reinforce the tunneling magnetoresistance and improve the sensitivity of resistance to low magnetic fields (compared to other types of layers). In [24] indicated that the magnetic field can easily cause the spin-flip in fairly thin layers of $Cr_2O_3$ in pressed $CrO_2$ powders. This can lead to changes in the properties of tunnel barriers. This question (as well as consideration of other possible causes of the positive MR) deserves further study.

**CONCLUSION**

The magnetoresistive effects in pressed powders consisting of $CrO_2$ ferromagnetic particles, separated by insulating layers is studied. It is shown that the resistance and the spin-dependent tunneling MR strongly depends on the particle shape (rounded or needle), as well as on the thickness and type of dielectric coating round the particles. These results indicate the possibility to influence the resistive characteristics of granular systems of $CrO_2$ particles of by varying the thickness and type of dielectric layers.

The paper presents new results of which should be particularly noted the following: (a) detection of non-monotonic dependence of MR with growing magnetic field (at low temperatures in powders with needle-shape particles the absolute value of MR at first grows rather rapidly with the field, and then begins to decrease markedly, forming maximum), (b) detection of anisotropy of MR, depending on the relative orientation of transport current and magnetic field, (c) description of a new method of synthesis that allows you to adjust the thickness of the dielectric shell.

**References**

1. J. M. D. Coey, *J. Appl. Phys.* **85**, 5576 (1999).
2. J. M. D. Coey and M. Venkatesan, *J. Appl. Phys.* **91**, 8345 (2002).
3. M. Ziese, *Rep. Progr. Phys.* **65**, 143 (2002).
4. J.M.D. Coey, A.E. Berkowitz, Ll. Balcells and F.F. Putris, *Phys. Rev. Lett.* **80**, 3815, (1998).
5. B. I. Belevtsev, N.V. Dalakova, M.G. Osmolowsky, E.Yu. Beliayev, A.A. Selutin, *Journal of Alloys and Compounds* **479**, 11 (2009).
6. E.Y. Tsymbal, O.N. Mryasov, P.R. LeClair, *J. Phys.: Condens. Matter* **15**, R109 (2003).
7. M.G. Osmolowsky, I.I. Kozhina, L.Yu. Ivanova, O.L. Baidakova, Zhurnal Prikladnoi Khimii (J. Appl. Chem.) **74**, 3 (2001).
8. B.I. Belevtsev, D.G. Naugle, K.D.D. Rathnayaka, A. Parasiris, J. Fink-Finowicki, *Physica B* **355**, 341 (2005).
9. A. G. Gamzatov. A. B. Batdalov, O. V. Melnikov, O. Yu. Gorbenko, Fizika Nizkikh Temperatur **35**, 290 (2009).
10. P. Sheng, *Phil. Mag. B* **65**, 357 (1992).
11. Sheng Ju, Tian-Yi Cai, and Z. Y. Li, *Appl. Phys. Lett,* **87,** 172504 (2005).
12. H. Kronmüller, in: F.C. Pu, Y.J.Wang, C.H. Shang (Eds.), *Aspects of Modern Magnetism*, World Scientific, Singapore, 1996, pp. 33–56.
13. I. Panagiotopoulos, N. Moutis, M. Ziese, A. Bollero, *J. Magn. Magn. Mater.* **299,** 94 (2006).
14. J. Smit, H.P.J. Wijn, *Ferrites,* Phillips Technical Library, Eindhoven, 1959.
15. H. Morrish, *Physical Principles of Magnetism*, John Wiley & Sons, New York,1965.
16. T. R. McGuire and R. I. Potter, *IEEE Trans. Magn.* **MAG-11**, 1018 (1975).
17. E. Dan Dahlberg, Kevin Riggs, and G. A. Prinz, J. *Appl. Phys.* **63**, 4270 (1988).
18. H. Liu, R.K. Zheng, Y. Wang, H.L. Bai, and X.X. Zhang, *Phys. Stat. Sol. (a)* **202,** 144 (2005).
19. H. Sun and Z. Y. Li, *Physics Letters A* **287**, 283 (2001).
20. S. Inoue, S. Maekawa, *Phys. Rev. B* **53**, R11927 (1996).
21. Chang He Shang, Janusz Nowak, Ronnie Jansen, and Jagadeesh S. Moodera, *Phys. Rev. B* **58,** R2917 (1998).
22. J. Soulen, J.H. Byers, M.S. Osofsky, B. Nadgorny, T. Ambrose, S.F. Cheng, P.R. Broussard, C.T. Tanaka, J. Nowak, J.S. Moodera, A. Barry, J.M.D. Coey, Science **282,** 85 (1998).
23. Jingping Wang, Ping Che, Jing Feng, Minfeng Lu, Jianfen Liu, Jian Meng, Yuanjia Hong,and Jinke Tang, , J. Appl. Phys. **97,** 073907 (2005).
24. R.K. Zheng, Hui Liu, Y. Wang, and X.X. Zhang, *Appl. Phys. Lett.* **84**, 702 (2004).

**Figure captions**

Figure 1. TEM micrograph of $CrO_2$ powder with needle-shaped particles. From this powder pressed samples №1, №2 and №3 were prepared, as described in the text.

Figure 2. Hysteresis curves of magnetization of the sample №3 at $T = 5.1$ K.

Figure 3. Temperature dependence of resistivity for the samples of $CrO_2$ pressed powders. Numbers of curves correspond to the sample's numbers.

Figure 4. MR hysteresis curves for the sample №4 at $T = 4.17$ K and $T = 20.43$ K. The insets show the behavior of MR in small magnetic fields. Arrows indicate the direction of the magnetic field sweeping for the curves recorded. The inset at T = 4.17 K shows the field $H_p$, at which the maximum value of resistance in a magnetic field is observed.

Figure 5. MR hysteresis curves for the sample №2 at different temperatures. Arrows indicate the direction of the magnetic field sweeping for the curves recorded.

Figure 6. The temperature dependences for the field $H_p$ (the peak of the positive MR) and the coercive force $H_c$ for the sample №3.

Figure 7. The temperature dependences of MR for the samples №2, №3 (in a field 1.5 T) and №4 (in the field 1.2 T).

Figure 8. MR hysteresis curves at $T = 4.4$ K for samples №1 (triangles) and №2 (squares). Arrows indicate the direction of the magnetic field sweeping for recording the curves.

**Table 1**
Characteristics of the powders of pressed $CrO_2$ investigated.

Legend: $S_{sp}$ - specific surface area; $H_c$ - the coercive force (at room temperature, and $T \approx 5$ K), $M_{rt}$ - specific magnetization at room temperature in the field $H = 1$ Tesla, $M_{max}$ - the maximum specific magnetization at low temperatures and high magnetic fields ($T \approx 5$ K, $H = 5$ T), MR - magnetoresistance, $[R(H) - R(0)]/R(0)$ at $T \approx 5$ K (the magnitude of the magnetic field is indicated in parentheses).

| Sample number | The surface shell of the particles | $S_{sp}$, м²g⁻¹ | $H_c$, T (room temp.) | $H_c$, T ($T \approx 5$ K) | $M_{rt}$, A·м²·kg⁻¹ | $M_{max}$, A·м²·кг⁻¹ | MR, % |
|---|---|---|---|---|---|---|---|
| 1 | Naturally degraded layer (mixture of $\beta$-CrOOH and chromic acid) | ~34 | 0,0429 | | 78,6 | | -19,98 (0,6 T) |
| 2 | $Cr_2O_3$ | ~34 | 0,0421 | | 72,5 | | -31,98 (0,4 T) |
| 3 | $Cr_2O_3$ | ~34 | 0,0422 | 0.0615 | 66,24 | 82,2 | -36,57 (0,4 T) |
| 4 | Stabilized layer $\beta$-CrOOH | 10,5 | 0,0149 | 0.033 | 62.5 | 88,3 | -18,65 (0,5 T) |

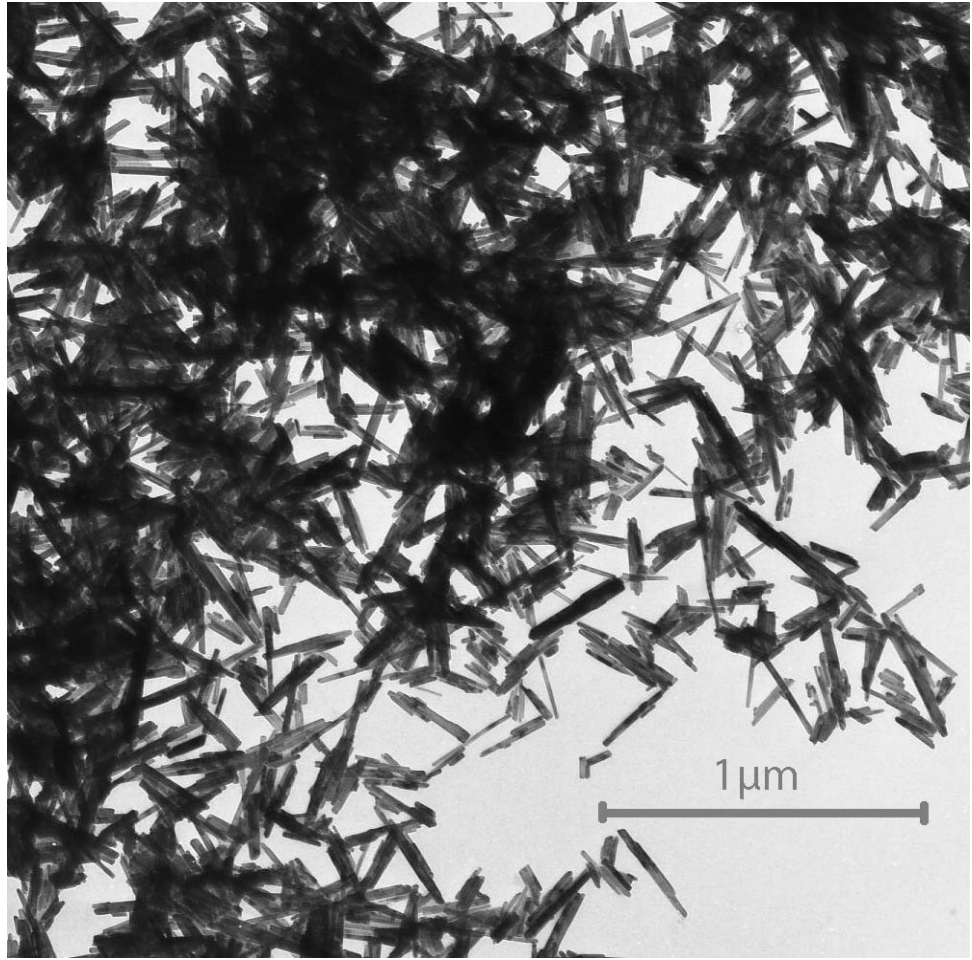

Fig.1

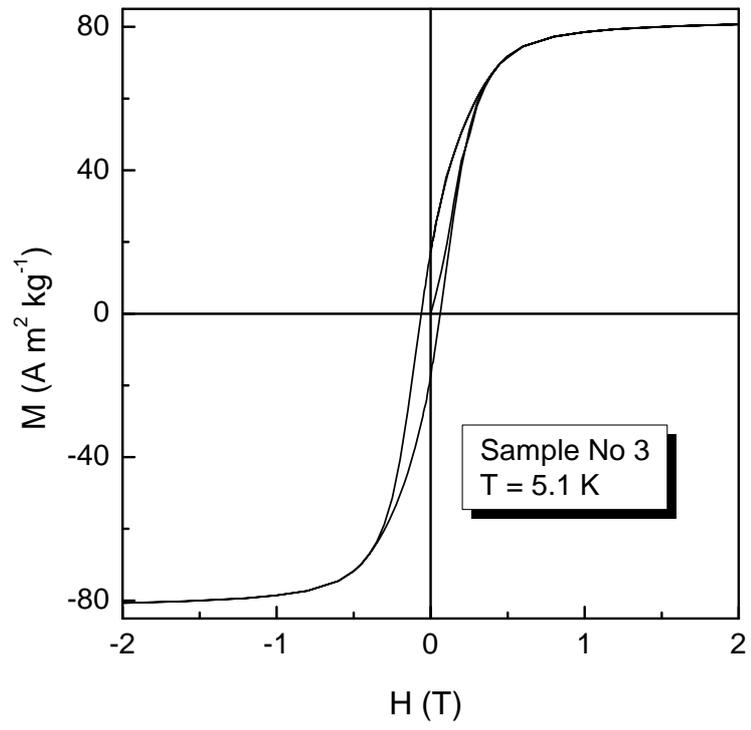

Fig.2

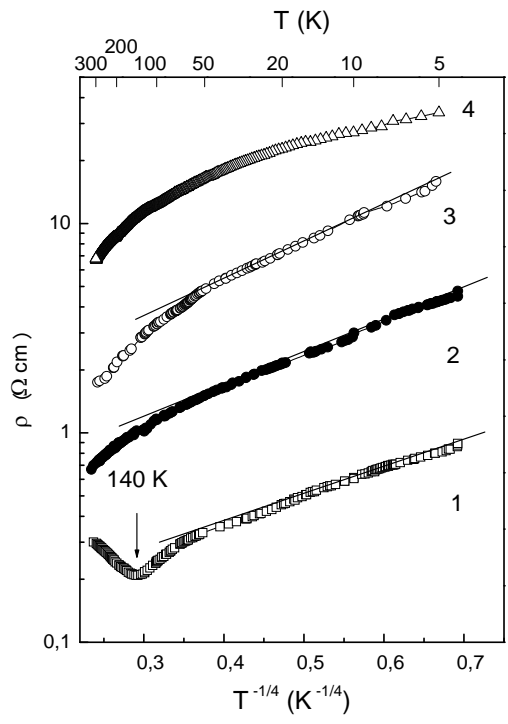

Fig.3

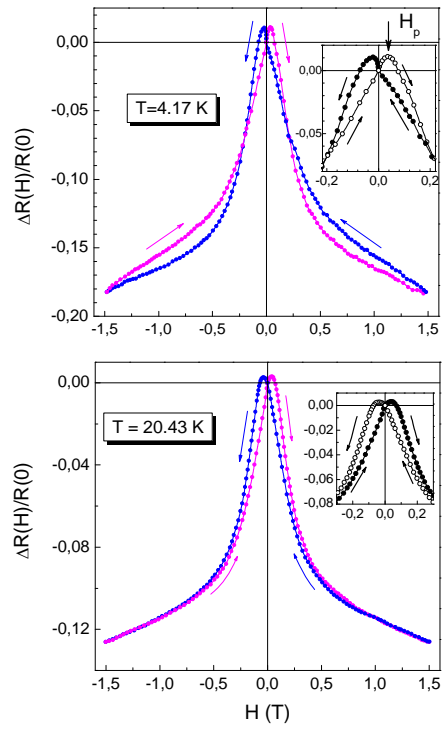

Fig.4

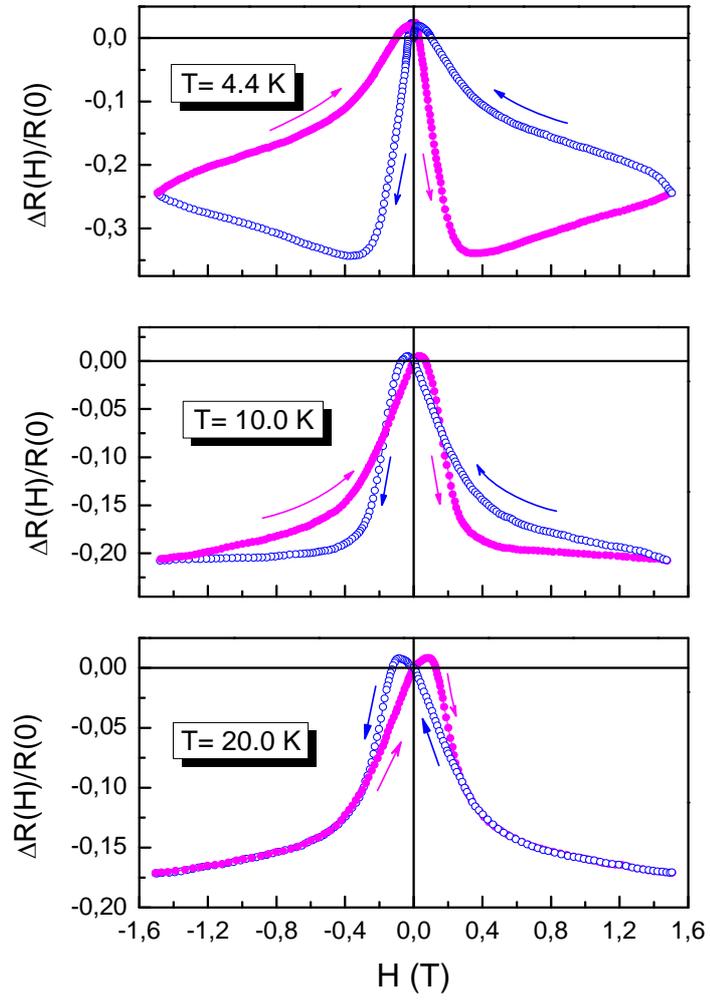

Fig.5

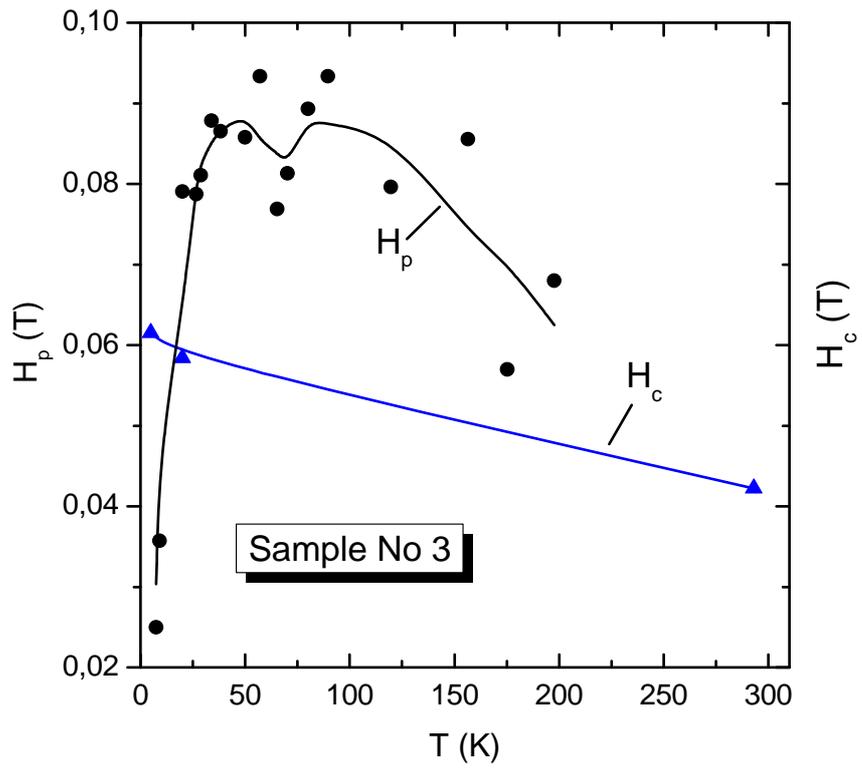

Fig.6

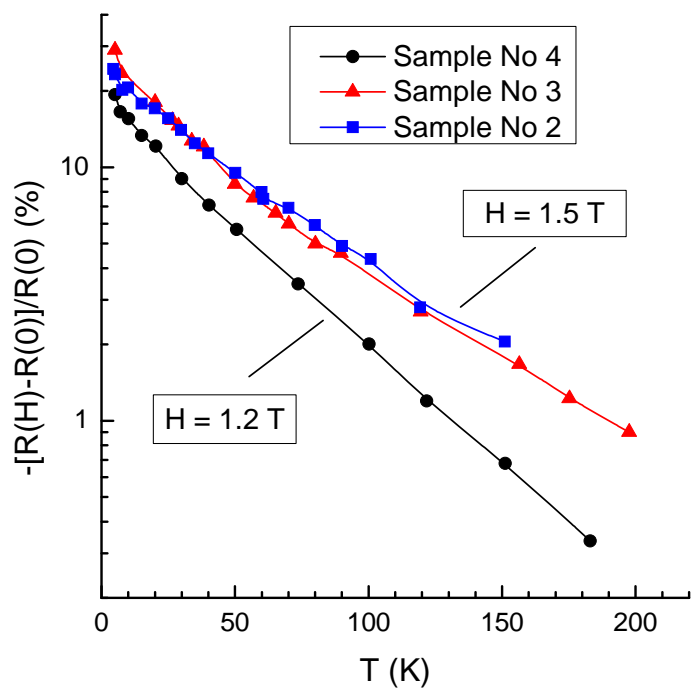

Fig.7

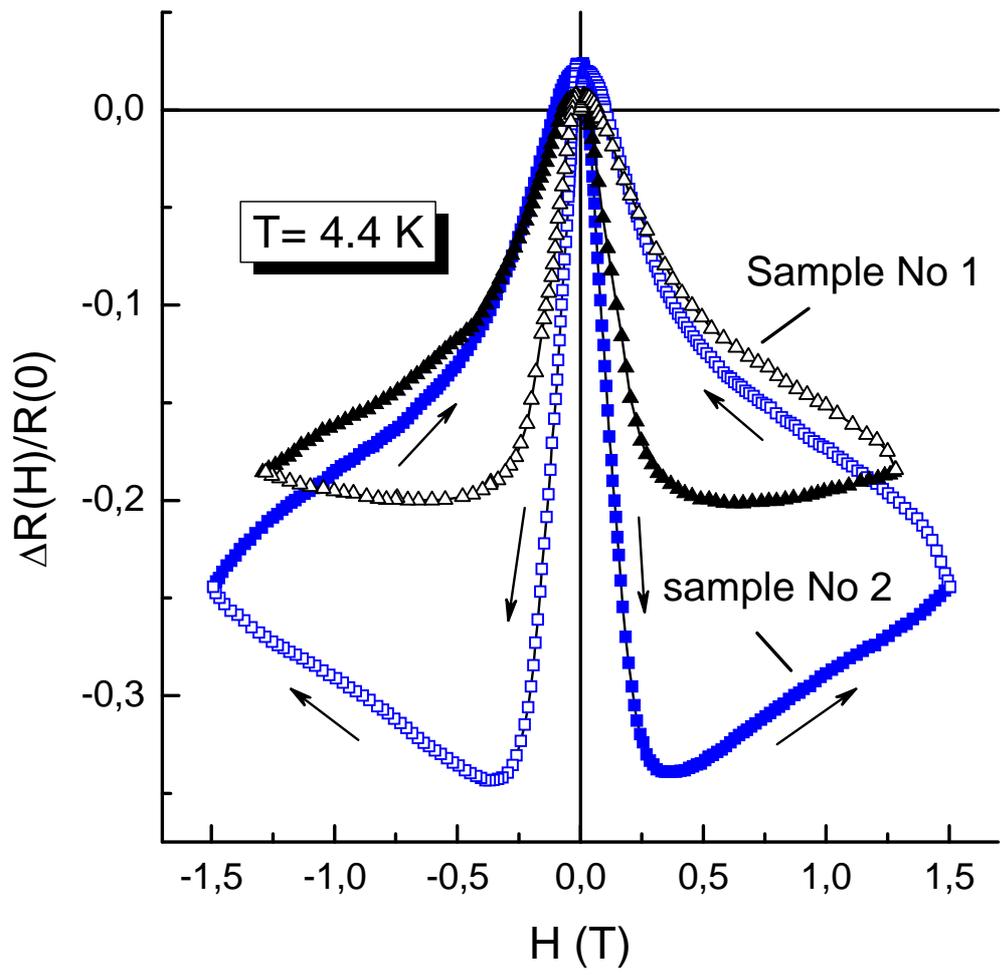

Fig.8